\documentclass[final,authoryear,10pt]{elsarticle}

\usepackage{amssymb}
\usepackage{amsthm}

\usepackage{amsmath}
\usepackage[latin1]{inputenc}

\journal{"Journal of Hydrology"}

\begin{document}

\begin{frontmatter}

\title{On the biases affecting water ages inferred from isotopic data}


\author[label1]{F.~J. Cornaton}
\author[label2]{Y.-J. Park}
\author[label3,label4]{E. Deleersnijder}

\address[label1]{Water Center for Latin America and the Caribbean,
 	Tecnol\'{o}gico de Monterrey,
 	Eugenio Garza Sada 2501 Sur, 64849 Monterrey, N.L., M\'{e}xico}
 	
\address[label2]{Department of Earth and Environmental Sciences,
	University of Waterloo,
	Waterloo, Ontario, Canada N2L 3G1}
 	
\address[label3]{Universit\'{e} Catholique de Louvain,
    Institute of Mechanics, Materials and Civil Engineering (IMMC),
	Avenue Georges Lema\^{\i}tre 4, B-1348 Louvain-la-Neuve, Belgium}
	
\address[label4]{Universit\'{e} Catholique de Louvain,
    Earth and Life Institute (ELI),
	Georges Lema\^{\i}tre Centre for Earth and Climate Research (TECLIM),
    Chemin du Cyclotron 2, B-1348 Louvain-la-Neuve, Belgium}

\begin{abstract}
Groundwater age has become a fundamental concept in groundwater hydrology, but ages originating from isotopic analyses are still identified with a lack of clarity and using models that occasionally are unrealistic. If the effect of advection and dispersion on water ages has already been extensively identified, very few studies address the reliability of using radiometric ages as derived from isotopic data to estimate aquifer properties such as average velocities. Using simple one-dimensional and two-dimensional analytical solutions for single-site and two-sites mobile-immobile systems, we compare the radiometric ages to the mean ages (or residence times) as deduced from a direct, physically-based simulation approach (using the mean age equation), and show that the competition between isotope decay rate and dispersion coefficient can generate important discrepancies between the two types of ages. A correction for the average apparent velocity originating from apparent isotopic ages is additionally provided. The particular case of the Tritium age dating method is also addressed, and a numerical example is finally given for illustrating the analysis considering a more complex and heterogeneous aquifer system. Our results suggest that age definitions based on the radioactivity of isotopes may not be representative for the mean age of the sample or for the groundwater velocity at given locations, and may not always be suitable for constraining the calibration of hydrogeological models.
\end{abstract}


\end{frontmatter}


\section{Introduction}
Radiometric age dating has been popular for decades in groundwater hydrology and for more than a century in other disciplines such as geology, archeology, and so on. It is one of the recent advances in hydrogeology to clarify the definitions of the kinematic, mean, and radiometric ages of groundwater. In general, there is an increasing consensus that the apparent radiometric age, which is the most measurable quantity, cannot be considered to represent the mean age, which may be the most representative for the average residence time of water molecules in the subsurface (\cite{Bethke2002},~\cite{Bethke2008},~\cite{Sanford2011}). Discrepancies between
ages obtained by means of different approaches have been frequently reported, and in the majority of the cases purely advective ages are compared to ages definitions that include advection and dispersion processes (\cite{Castro2005},~\cite{LaBolle2006},~\cite{Bethke2008}). Concentration-based ages are simply apparent ages that do not a priori relate to the mean age of a water sample (\cite{Sanford2011}). Radiometric ages are classically derived from the interpretation of chemical/isotopic data that ignores transport characteristics although isotope transport undergoes dispersion/diffusion, decay and to some extend reaction processes in the subsurface. The kinematic age of groundwater is defined as the time for which a groundwater molecule may travel along a one-dimensional stream line from the time of recharge (birth). Thus, this kinematic age is calculated based only on the advective component of the migration of water molecules (piston flow) without considering the mixing processes among water molecules which enter the system at different times in the past (different birthdates). Since the kinematic age does not consider mixing, it cannot be directly translated from any chemical signature in the real world where various types of mixing always exist. Irrespective of the importance of the kinematic age it should be noted that it is a function of only the groundwater velocity which is one of the most fundamental physical parameters to describe the groundwater flow system. When the ages are known at two different locations, hydrogeologists may hope to be able to calculate the average velocity of groundwater dividing the distance by the difference in age along this distance.\\
A radiometric age is a measurable quantity for a given water sample, based on the change in the activity of decaying solutes. When the initial activity at birth and the rate of the decay (change) are known, the radiometric age can be calculated from the current activity at given locations. This assumes that the water sample under consideration is in a closed system, i.e. a system that does not exchange any material with its surrounding. Radiometric age thus differs from the mean age of groundwater, which is defined as the first temporal moment of the age probability density function for given samples, describing how age distributes as a result of advection and mixing processes. Age mass is a useful extensive (additive) property, defined as the product of age and mass for a given mass (e.g. see~\cite{Goode1996},~\cite{Varni1998},~\cite{Ginn1999}). Mathematically an intensive property or the age probability density, defined as the age mass divided by the total mass, is a popular concept to derive the equation that describes the mixture of different ages of groundwater.\\
\cite{Bethke2008} clarify the difference between the sample apparent (radiometric) age and the mean age of groundwater, the mathematical description of which has been well established by~\cite{Goode1996},~\cite{Varni1998},~\cite{Ginn1999},~\cite{Cornaton2006}, and~\cite{Ginn2009} among others. Although it seems that there are many different versions of mathematical equations to describe the mean age, age probability density, and age mass, conceptual difference is minor and consists in the assumed steady or transient flow conditions or different dependent variables as an integral moment or probability density (\cite{Ginn2007},~\cite{Cornaton2007}).\\
The recent finding that the radiometric age may not be representative for the mean residence time or the velocity of groundwater can imply that the only way to analyze the mean residence time is to consider applications of the mean age equation. This may lead to questioning the relevance of the radiometric ages frequently estimated by isotope hydrologists and for what purpose the radiometric age is measured. Such an age may not be the most relevant data to be used for calibrating a groundwater flow field in a numerical model (for example by solving inverse problems constrained by radiometric age data and traditional hydrological data). As indicated by~\cite{Bethke2008}, time is ripe for developing a new paradigm that would account for the recent findings.\\
In this paper, we first review mathematical equations to describe the transport of the kinematic age, mean age, and decaying solute and illustrate the differences among them. The relationships between a measurable quantity and more intrinsic quantities are then analyzed. A discussion on what can be interpreted from measurable quantities and under which conditions is finally made.

\section{Mathematical Description of Kinematic, Radiometric and Mean Ages}
This Section is dedicated to the definition of the kinematic, radiometric and mean ages, and to the comparison between each age property. One-dimensional and two-dimensional analyses are made in order to reveal the bias induced by the use of apparent ages for calculating apparent flow velocities.

\subsection{One-Dimensional Analysis}
Considering a simple one-dimensional flow system where groundwater recharges at the origin ($x = 0$) and the flow system is homogeneous in a steady state regime ($v(x,t)=v$), the kinematic age ($A_k$) is straightforward to calculate from its definition:
\begin{equation}
A_k = \frac{x}{v}
\label{eq:kinematicAge}
\end{equation}
Steady-state transport of a decaying solute can be described using a simple advection-dispersion-reaction equation as:
\begin{equation}
- v \frac{{d C(x)}}{{d x}} + D \frac{{d^2 C(x)}}{{d x^2 }} - \lambda C(x) = 0
\label{eq:1DADE}
\end{equation}
where $D$ is the dispersion coefficient and $\lambda$ is the constant for the first order decay to satisfy $C(t) = C(t = 0)e^{- \lambda t}$. Since the boundary concentration at origin is assumed to be known, $C(x = 0) = C_0$, the solution of Eq.~(\ref{eq:1DADE}) is
\begin{equation}
C(x) = C_0 \exp \left ( \frac{x v (1-\beta)}{2 D} \right )
\label{eq:concSol}
\end{equation}
where
\begin{equation}\label{eq:betadef}
\beta = \sqrt{1 + \frac{4 \lambda D}{v^2}}
\end{equation}
and where use has also been made of the condition $- \left.D \frac{\partial C(x)}{\partial x} \right|_{x = \infty } = 0$. The apparent age ($A_a$) is a direct translation of the initial concentration and the rate of change in the first order kinetic decay and is classically given as
\begin{equation}
A_a(x) = - \frac{1}{\lambda} \ln \left( \frac{C}{C_0} \right) = \frac{x v}{2\lambda D}\left( \beta - 1 \right)
\label{eq:apparentAge}
\end{equation}
It is obvious from Eq.~(\ref{eq:apparentAge}) that the apparent age is influenced by the mixing ($D$) and the solute decay rate ($\lambda$) as well as by the flow velocity ($v$) and the location ($x$). Eqs.~(\ref{eq:kinematicAge}) and~(\ref{eq:apparentAge}) also indicates that apparent age $A_a$ can only relate to kinematic age for the asymptotic situation, i.e. $A_a = A_k$ when $\frac{v^2}{\lambda D} \rightarrow \infty$. \\

The equation describing mean age ($A_m$) is well established (e.g. see~\cite{Goode1996},~\cite{Varni1998}) and a simplified version for the given one-dimensional domain can be written as
\begin{equation}
- v \frac{d A_m(x)}{dx} + D\frac{d^2 A_m(x)}{dx^2} + 1 = 0
\label{eq:meanAgeEqn}
\end{equation}
When the birth for groundwater is assumed at the inlet and a boundary condition for the mean age is given as $A_m(x = 0) = 0$, the mean groundwater age is same as the kinematic age:
\begin{equation}
A_m(x) = \frac{x}{v}
\label{eq:meanAgeSol}
\end{equation}
for which the condition $- \left.D \frac{\partial A_m(x)}{\partial x} \right|_{x = \infty } = 0$ has also been used. Equations~(\ref{eq:kinematicAge}) and~(\ref{eq:meanAgeSol}) imply that if the flow system is one-dimensional or if there is only one velocity component, the mean age which is same as the kinematic age is only a function of a flow velocity and of the location in the flow system, and thus it is straightforward to translate the mean age to describe the flow velocity. In other words, when one knows the groundwater mean ages ($A_m(x_1)$ and $A_m(x_2)$) at two different locations ($x_1$ and $x_2$) in a one-dimensional flow system, the distance between two locations divided by the difference in mean age can provide an average velocity between two points:
\begin{equation}
v = \frac{x_2 - x_1}{A_m(x_2) - A_m(x_1)}
\label{eq:veloApproxFromMA}
\end{equation}
When the apparent radiometric ages ($A_a(x_1)$ and $A_a(x_2)$) are known, the average velocity between two locations can also be calculated from~(\ref{eq:apparentAge}) such that
\begin{equation}
v = v_a - \frac{\lambda D}{v_a}
\label{eq:veloApproxFromAA}
\end{equation}
where the apparent velocity $v_a$ is defined as $[x_2  - x_1 ]/[A_a (x_2 ) - A_a (x_1 )]$, by analogy with Eq.~(\ref{eq:veloApproxFromMA}). Eq.~(\ref{eq:veloApproxFromAA}) indicates that the groundwater velocity can be calculated using the apparent velocity when the decaying property of the solute ($\lambda$) is known and the mixing processes ($D$) are clearly defined. It is also implied from~(\ref{eq:veloApproxFromAA}) that the apparent velocity generally overestimates the actual velocity and that the difference between them becomes greater when $v_a$ is smaller and the product $\lambda D$ is larger. When the dispersion coefficient is defined as $D = D_{diff} + \alpha_l v$, Eq.~(\ref{eq:veloApproxFromAA}) can be re-arranged to give
\begin{equation}
v = \frac{v_a (1 - \lambda D_{diff} /v_a^2)}{1 + \lambda \alpha_l / v_a}
\label{eq:veloApproxFromAA2}
\end{equation}
where $D_{diff}$ is the effective diffusion coefficient in the porous medium and $\alpha_l$ is the coefficient of longitudinal dispersivity. Defining the dimensionless variable $\Pi_a = \frac{v_a^2}{\lambda D}$ (defined as Jenkins number in~\cite{Mouchet2008}), Eq.~(\ref{eq:veloApproxFromAA}) takes the form:
\begin{equation}
v = v_a \left (1 - \frac{1}{\Pi_a} \right )
\label{eq:veloApproxFromAAadim}
\end{equation}
The variable $\Pi_a$ is similar to the P\'{e}clet number ($Pe$) if one considers a characteristic distance $L = \frac{v_a}{\lambda}$ as being the distance that tracer molecules can travel if they move at the average velocity $v_a$ during time $\lambda^{-1}$ ($Pe = \frac{v_a L}{D} = \frac{v_a^2}{\lambda D} = \Pi_a$). \\

Finally, the mean age of a radioactive tracer ($A_{m,\lambda}$) can be obtained by calculating the first two temporal moments of the corresponding age concentration density function (pdf). The age concentration pdf $g_\lambda(x,\tau)$ of a radioactive tracer (where $\tau$ denotes the variable age) is solution of the advection-dispersion-reaction equation
\begin{equation}\label{eq:tracerpdfEqn}
\frac{\partial g_\lambda(x,\tau)}{\partial \tau} = - v \frac{\partial g_\lambda(x,\tau)}{\partial
x} + D \frac{\partial^2 g_\lambda(x,\tau)}{\partial x^2} - \lambda g_\lambda(x,\tau)
\end{equation}
with the boundary conditions $g_\lambda(0,\tau) = \delta(\tau)$ (zero age flux concentration) and $- \left.D \frac{\partial g_\lambda(x,\tau)}{\partial x} \right|_{x = \infty } = 0$:
\begin{equation}\label{eq:tracerpdfSol}
g_\lambda(x,\tau) = \frac{x}{\sqrt{4 \pi D} \tau^{3/2}} \exp \left ( -\frac{( x - v \tau )^2}{4 D \tau} - \lambda \tau \right ) = \exp \left ( -\lambda \tau \right ) g_0(x,\tau)
\end{equation}
For instance, the zero-order temporal moment of $g_\lambda(x,\tau)$ gives the mass of radioactive tracer in the system:
\begin{equation}\label{eq:componentM0}
m_{0,\lambda}(x) = \int_0^\infty {g_\lambda(x,\tau) d\tau} = \exp \left ( \frac{x v (1-\beta)}{2 D} \right )
\end{equation}
in which $\beta$ has been defined in Eq.~(\ref{eq:betadef}). The first-order temporal moment of $g_\lambda(x,\tau)$ gives the mean age concentration of the radioactive tracer:
\begin{equation}\label{eq:componentM1}
m_{1,\lambda}(x) = \int_0^\infty {\tau g_\lambda(x,\tau) d\tau} = \frac{x}{\beta v} m_{0,\lambda}(x)
\end{equation}
The mean age of the radioactive tracer is then obtained as
\begin{equation}\label{eq:componentMA}
A_{m,\lambda}(x) = \frac{m_{1,\lambda}(x)}{m_{0,\lambda}(x)} = \frac{x}{\beta v} = \frac{A_m(x)}{\beta}
\end{equation}
Note that $m_{0,\lambda}(x)$ satisfies the tracer mass equation~(\ref{eq:1DADE}) using the boundary conditions $m_{0,\lambda}(0) = 1$ (unit mass input) and $- \left.D \frac{\partial m_{0,\lambda}(x)}{\partial x} \right|_{x = \infty } = 0$, and that $m_{1,\lambda}(x)$ satisfies the tracer mean age concentration equation
\begin{equation}\label{eq:componentM1Eqn}
- v \frac{\partial m_{1,\lambda}(x)}{\partial
x} + D \frac{\partial^2 m_{1,\lambda}(x)}{\partial x^2} - \lambda m_{1,\lambda}(x) + m_{0,\lambda}(x) = 0
\end{equation}
with the boundary conditions $m_{1,\lambda}(0) = 0$ (zero mean age concentration) and $- \left.D \frac{\partial m_{1,\lambda}(x)}{\partial x} \right|_{x = \infty } = 0$.\\

\begin{figure}[h]
  \begin{center}
  \includegraphics[width=\textwidth]{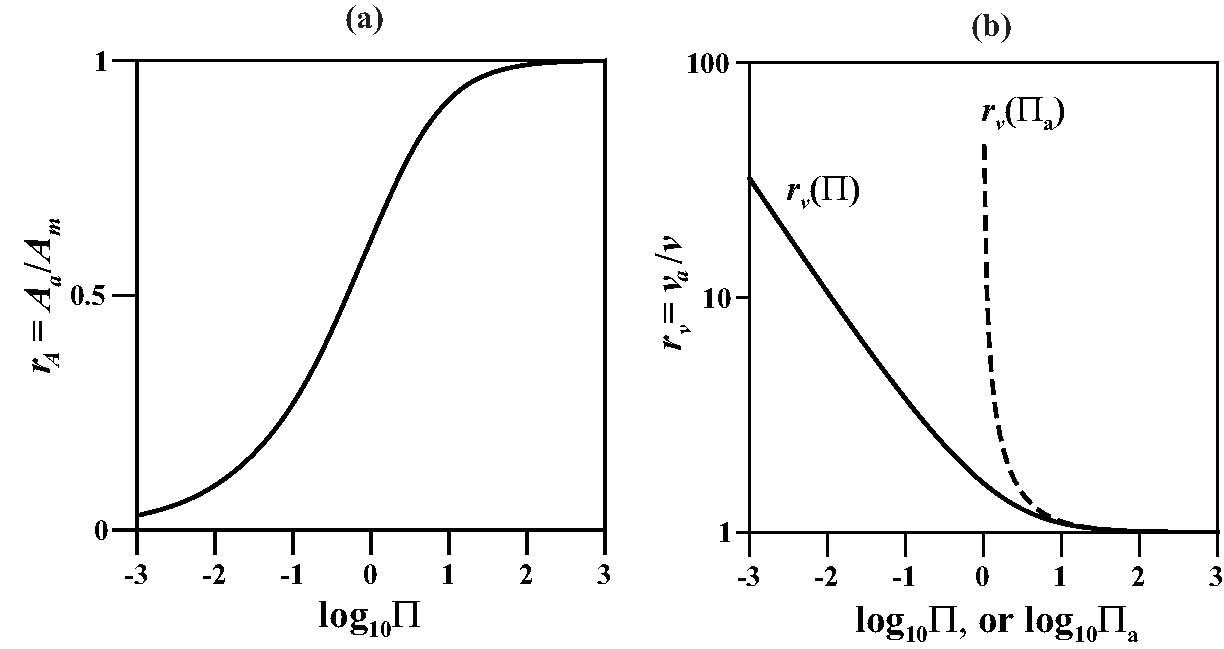}
  \end{center}
  \caption{\small The ratios of (a) apparent and mean ages, and (b) apparent and mean velocities as functions of the dimensionless variables $\Pi = v^2 /\lambda D$ and $\Pi_a = v_a^2 /\lambda D$.}
 \label{fig:fig1}
\end{figure}

Dimensionless variables can be used to describe the relations between the apparent age, mean tracer age and mean water age, and between the apparent velocity and the actual velocity. The ratio of the apparent and mean ages $r_A = A_a /A_m$ and the ratio of the apparent and mean velocities $r_v = v_a / v$ can be defined using the dimensionless variables $\Pi = v^2 /\lambda D$, $\Pi_a = v_a^2 /\lambda D$, $\Pi_{a,diff} = v_a^2 /\lambda D_{diff}$, and $\Pi_{a,disp} = v_a /\lambda \alpha _l$. The following relations are derived from~(\ref{eq:apparentAge}),~(\ref{eq:veloApproxFromAA}), and~(\ref{eq:veloApproxFromAA2}) (see Fig.~\ref{fig:fig1}):
\begin{subequations}\label{eq:ratios}
\begin{equation}\label{eq:ratios1}
r_A = \frac{\Pi }{2}\left( {\sqrt {1 + \frac{4}{\Pi}} - 1} \right)
\end{equation}
\begin{equation}\label{eq:ratios2}
r_v = \frac{{\Pi_a }}{{\Pi_a - 1}} = \frac{1}{2}\left( {1 + \sqrt {1 + \frac{4}{\Pi}}} \right) = \frac{{1 + 1/\Pi_{a,disp} }}{{1 - 1/\Pi_{a,diff}}}
\end{equation}
\end{subequations}
A first fundamental consideration arising from the analysis of Eqs.~(\ref{eq:componentMA}),~(\ref{eq:meanAgeSol}) and~(\ref{eq:ratios1}) is that the inequality $A_{m,\lambda} \leq A_a \leq A_m$ always holds, as already shown by~\cite{Deleersnijder2001}. \cite{Delhez2003} also demonstrated that this inequality is valid for more general contexts. Another interesting aspect is that the age and velocity ratios in Eqs.~(\ref{eq:ratios1}) and~(\ref{eq:ratios2}) are independent on the $x$-coordinate. Eq.~(\ref{eq:ratios1}) and Fig.~\ref{fig:fig1}a indicate that the apparent age generally underestimates the mean age and that the ratio between two ages becomes larger as the decay and dispersion dominate the advective transport and vice versa. Eq.~(\ref{eq:ratios1}) also indicates that the age ratio $r_A$ does not depend on the spatial coordinate $x$. When the parameter $\Pi$ decreases (i.e. $\lambda D$ increases), then the bias increases, and vice-versa. Thus apparent ages are reliable only if $\Pi$ is large enough, or consequently only if the product $\lambda D$ is small enough. In Eq.~(\ref{eq:ratios2}) and Fig.~\ref{fig:fig1}b, the apparent velocity is similar to the actual velocity when the transport regime is advection-dominated ($\Pi > 1$), and the actual velocity can be calculated from the apparent velocity when $\Pi_a > 1$. When the dispersion dominates the transport ($\Pi_a \leq 1$), it may not be possible to estimate the actual velocity from the apparent velocity.

\subsection{Coupled One-Dimensional Solutions for Two-Dimensional Analysis of Aquifer-Aquitard Systems}
\begin{figure}[h]
  \begin{center}
  \includegraphics[width=0.6\textwidth]{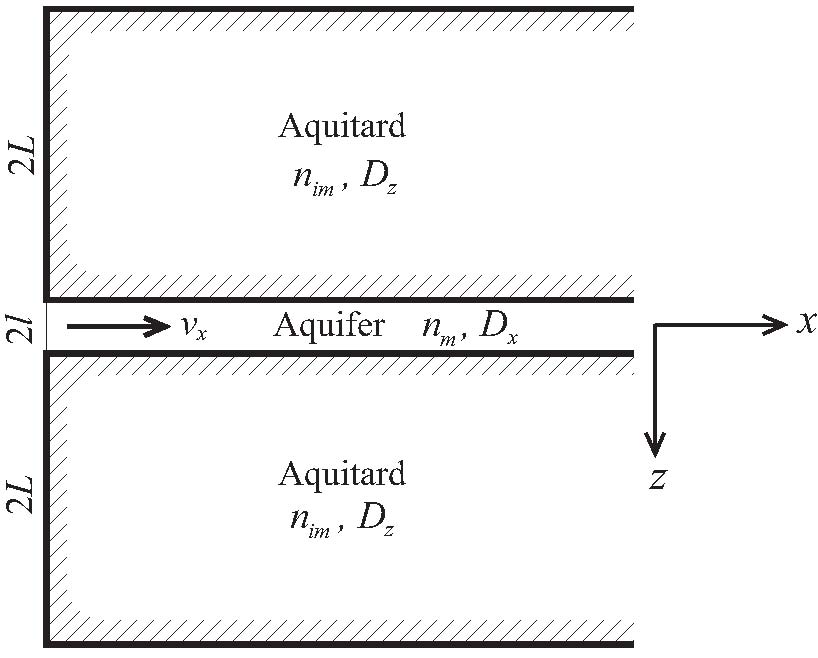}
  \end{center}
  \caption{\small Illustration of the aquifer-aquitard system used to derive two-dimensional analytical solutions.}
 \label{fig:fig2}
\end{figure}

A simple two-dimensional domain can be considered when mobile and immobile (or a lot less mobile) waters co-exist in a groundwater system. Such two-dimensional analyses have been widely adapted for a fracture-matrix system (\cite{Cornaton2008}) and also in an aquifer-aquitard system (\cite{Bethke2002,Bethke2008},~\cite{Ginn2009}). In these flow systems, the flow in a mobile region is essentially one-dimensional and the velocity is large compared to that in an immobile area so that it is safe to assume that the velocity in the immobile region is negligible. When an aquitard (or a matrix block) has a dimension of $2L$ perpendicular to the flow direction $x$ between two aquifers of a dimension $2l$ (or two fractures of an aperture value of $2l$), transport of the mean age or decaying solute can be described using an advection-dispersion-reaction equation in the aquifer and the diffusion-reaction equation in the aquitard (see Figure~\ref{fig:fig2}).\\

The kinematic age in the mobile aquifer ($A_{k,m}$) is, by definition (see Eq.~(\ref{eq:kinematicAge})), a direct translation of the velocity in the mobile region ($v_x$):
\begin{equation}
A_{k,m} = \frac{x}{v_x}
\label{eq:mobilekinematicAge}
\end{equation}
Since the velocity in the immobile region is assumed to be negligible, the kinematic age in the immobile aquitard ($A_{k,im}$) is infinite for all $x$ and $z$, where $z$ represents a direction perpendicular to the flow direction along the aquifer.\\

\begin{figure}[h]
  \begin{center}
  \includegraphics[width=\textwidth]{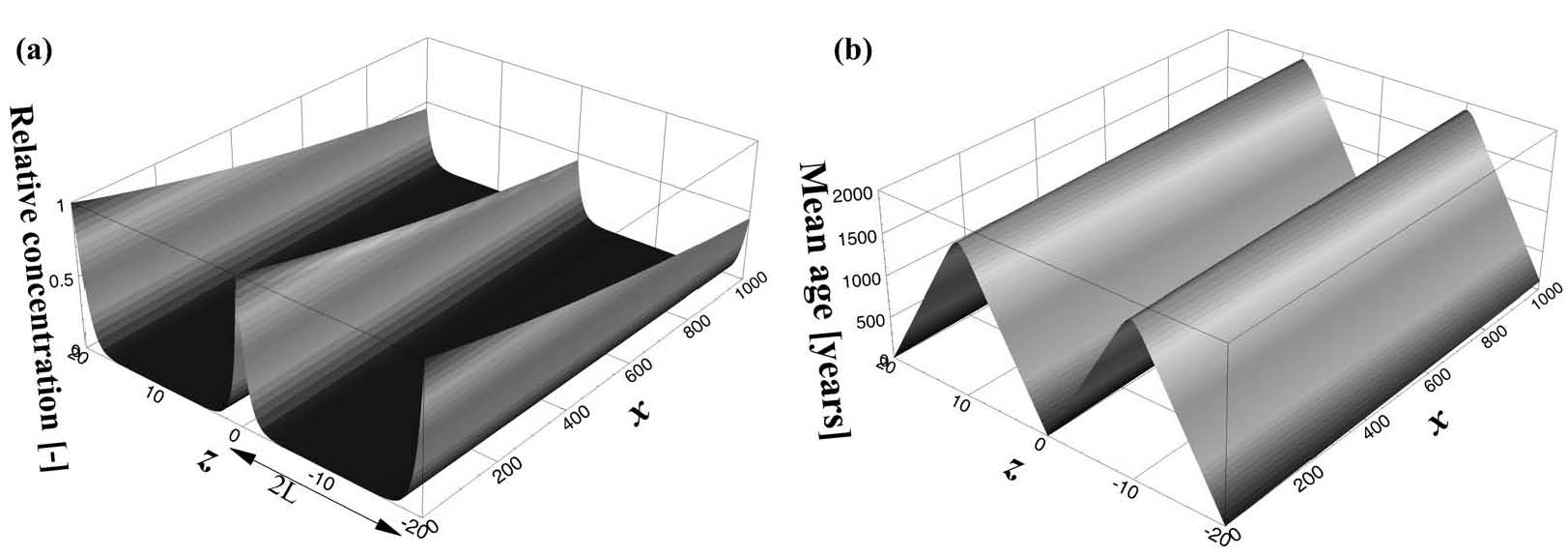}
  \end{center}
  \caption{\small Illustration of the distributions of (a) relative concentration of a decaying solute and (b) apparent age in a two-dimensional aquifer-aquitard system.}
 \label{fig:fig3}
\end{figure}

Steady-state transport of a decaying solute in the aquifer is described by the following one-dimensional advection-dispersion-reaction equation such that
\begin{equation}
- v_x \frac{d C_m(x)}{dx} + D_x \frac{d^2 C_m(x)}{dx^2} - \lambda C_m(x) + \left. {\frac{n_{im} D_z }{n_m l}\frac{d C_{im}(z|x)}{dz}} \right|_{z = 0}  = 0
\label{eq:mobileADE}
\end{equation}
where $n_m$ and $n_{im}$ represent the porosities in the mobile and immobile regions, respectively. Steady-state transport in the aquitard can also be described by a one-dimensional diffusion-reaction equation:
\begin{equation}
D_z \frac{d^2 C_{im} (z|x)}{dz^2} - \lambda C_{im}(z|x) = 0
\label{eq:immobileADE}
\end{equation}
When the concentrations in the interface between mobile and immobile regions are assumed to be identical ($C_{im} (z|x) = C_m (x)$) and the mass flux in the middle of the aquitard is negligible ($\left. { - D_z \frac{d C_{im}(z|x)}{dz}} \right|_{z = L}  = 0$), then the solution of Eq.~(\ref{eq:immobileADE}) is given as
\begin{equation}
C_{im}(z|x) = C_m(x)\frac{{e^{\sqrt {{\textstyle{\lambda \over {D_z}}}} (z - L)} + e^{- \sqrt {{\textstyle{\lambda \over {D_z}}}} (z - L)}}}{{e^{\sqrt {{\textstyle{\lambda \over {D_z}}}} L}  + e^{- \sqrt {{\textstyle{\lambda \over {D_z}}}}L}}}
\label{eq:immobileSol}
\end{equation}
Using Eq.~(\ref{eq:immobileSol}), Eq.~(\ref{eq:mobileADE}) can be re-written as
\begin{equation}
- v_x \frac{d C_m(x)}{dx} + D_x \frac{d^2 C_m(x)}{dx^2} - (\lambda + \gamma) C_m(x) = 0
\label{eq:mobileADE2}
\end{equation}
where
\begin{equation}
\gamma = \frac{{n_{im} \sqrt {\lambda D_z }}}{{n_m l}}\frac{{e^{\sqrt {{\textstyle{\lambda \over {D_z }}}} L} - e^{- \sqrt {{\textstyle{\lambda \over {D_z }}}} L} }}{{e^{\sqrt {{\textstyle{\lambda \over {D_z }}}} L} + e^{- \sqrt {{\textstyle{\lambda \over {D_z}}}} L} }}
\label{eq:gamma}
\end{equation}
The solution of Eq.~(\ref{eq:mobileADE2}) is similar to Eq.~(\ref{eq:concSol}) where $\lambda$ is replaced by $(\lambda + \gamma)$:
\begin{equation}
C_m(x) = C_m(0) \exp \left[ {\frac{x v_x}{2 D_x}\left( {1 - \sqrt {1 + \frac{{4(\lambda + \gamma)D_x}}{{v_x^2}}}} \right)} \right]
\label{eq:mobileConc}
\end{equation}
and the apparent age in the aquifer ($A_{a,m}$) is also similar to Eq.~(\ref{eq:apparentAge}) such that
\begin{equation}
A_{a,m}(x) = - \frac{1}{\lambda} \ln \left( {\frac{C_m(x)}{C_m(0)}} \right) = \frac{x v_x}{2\lambda D_x}\left( {\sqrt {1 + \frac{4(\lambda + \gamma)D_x}{v_x^2}} - 1} \right)
\label{eq:mobileAA}
\end{equation}
The apparent age in the aquitard ($A_{a,im}$) can also be calculated using~(\ref{eq:immobileSol}) and~(\ref{eq:mobileConc}):
\begin{equation}
A_{a,im}(x,z) = - \frac{1}{\lambda} \ln \left( {\frac{C_{im}(z|x)}{C_m(0)}} \right) = A_{a,m}(x) + \frac{1}{\lambda} \ln \left[ {\frac{{e^{\sqrt {{\textstyle{\lambda \over {D_z}}}} L} + e^{- \sqrt {{\textstyle{\lambda \over {D_z}}}}L}}}{{e^{\sqrt {{\textstyle{\lambda \over {D_z}}}} (z - L)} + e^{- \sqrt {{\textstyle{\lambda  \over {D_z}}}} (z - L)}}}} \right]
\label{eq:immobileAA}
\end{equation}
The results~(\ref{eq:mobileAA}) and~(\ref{eq:immobileAA}) indicate that the apparent age in the aquifer is determined by the decay rate of the solute and the mixing processes in both the aquifer and aquitard, and that the apparent age in the aquitard is same as that in the aquifer at the interface between the two domains and has a maximum value at the center of the aquitard when $z = L$. The maximum of the apparent age is found at $z = L$:
\begin{equation}
\max [A_{a,im}(x,z)] = A_{a,im}(x,z = L) = A_{a,m} + \frac{1}{\lambda} \ln \left[ {\frac{{e^{\sqrt {{\textstyle{\lambda \over {D_z}}}} L} + e^{- \sqrt {{\textstyle{\lambda \over {D_z}}}} L}}}{2}} \right]
\label{eq:MaximmobileAA}
\end{equation}
The maximum age increases as the aquitard becomes thicker (larger $L$), the decaying rate of the solute is higher (larger $\lambda$), and the effective diffusion coefficient ($D_z$) is smaller. Fig.~\ref{fig:fig3} illustrates the distributions of the concentration of a decaying solute and the apparent age in such a two-dimensional system when $l = 5$m, $L = 10$m, $v_x = 10$m/yr, $\lambda = \ln 2/100{\rm{yr}}$, $D_x = \alpha_l v_x  = 100{\rm{m}}^{\rm{2}} {\rm{/yr}}$, $D_z = 0.004{\rm{m}}^{\rm{2}} {\rm{/yr}}$, and $n_m = n_{im} = 0.1$.\\

\begin{figure}[h]
  \begin{center}
  \includegraphics[width=0.90\textwidth]{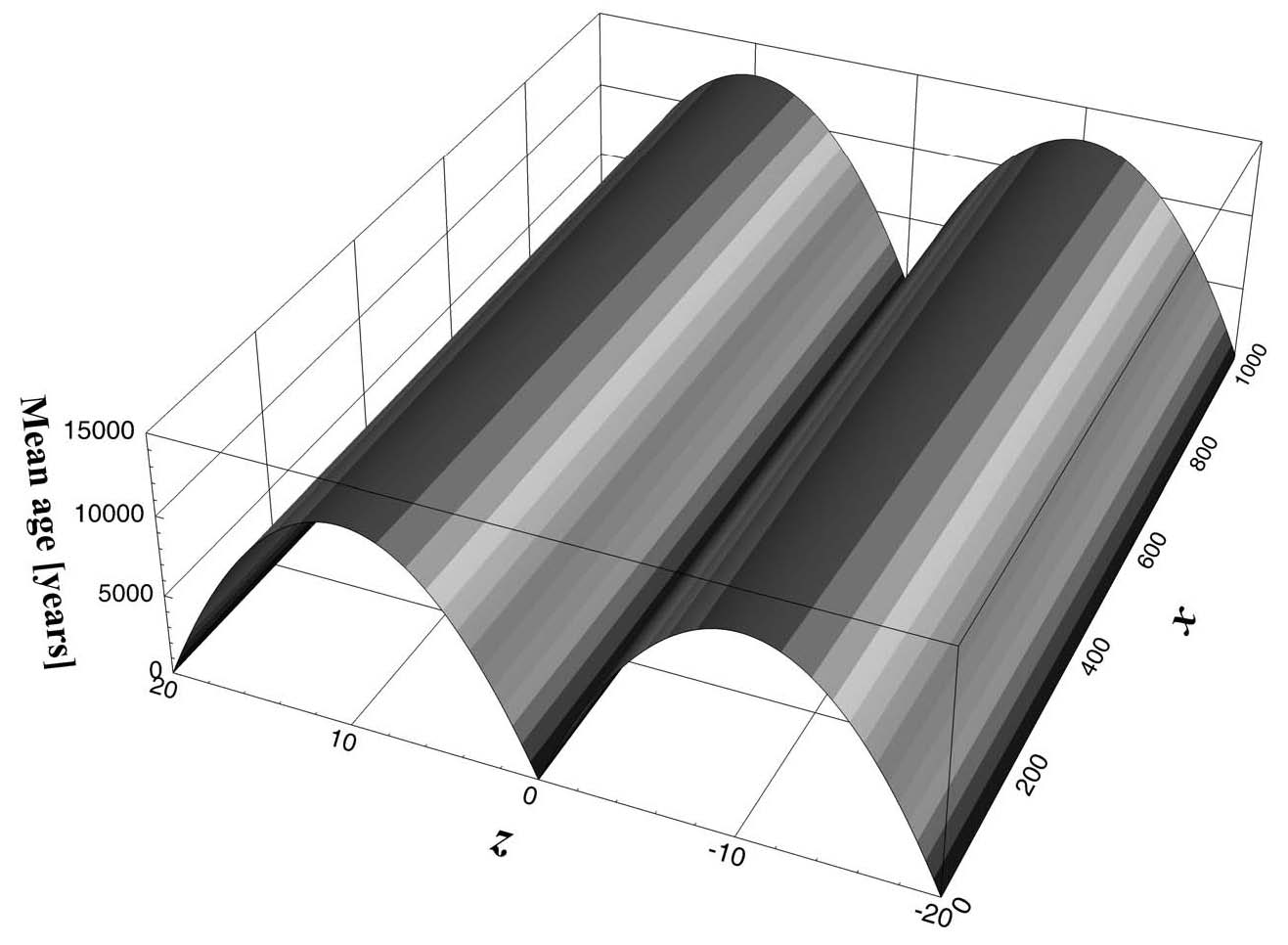}
  \end{center}
  \caption{\small Illustration of the groundwater mean age distribution in a two-dimensional aquifer-aquitard system.}
 \label{fig:fig4}
\end{figure}

\begin{figure}[h]
  \begin{center}
  \includegraphics[width=\textwidth]{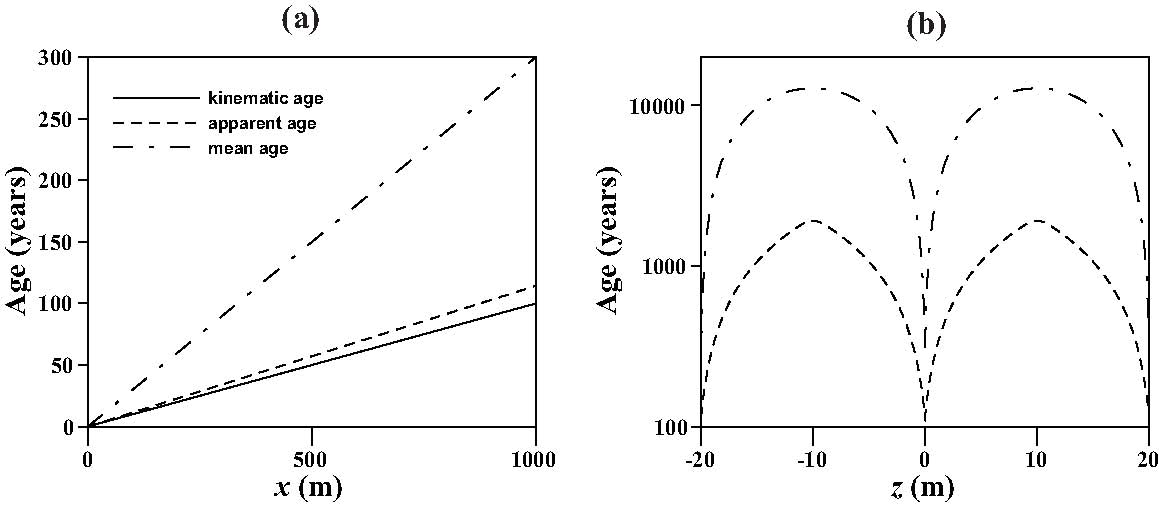}
  \end{center}
  \caption{\small Illustration of the kinematic, apparent, and groundwater mean age distributions (a) along the flow direction in the mobile region and (b) across the immobile region ($x = 1000$m) of the two-dimensional system illustrated in Figures~(\ref{fig:fig3}) and~(\ref{fig:fig4}).}
 \label{fig:fig5}
\end{figure}

Mean groundwater age in the aquifer ($A_{m,m}$) and in the aquitard ($A_{m,im}$) can be described using equations similar to~(\ref{eq:meanAgeEqn}),~(\ref{eq:mobileADE}) and~(\ref{eq:immobileADE}):
\begin{equation}
- v_x \frac{d A_{m,m}(x)}{dx} + D_x \frac{d^2 A_{m,m}(x)}{dx^2} + \left. {\frac{n_{im} D_z}{n_m l}\frac{d A_{m,im}(z|x)}{dz}} \right|_{z = 0} + 1 = 0
\label{eq:mobileMAEq}
\end{equation}
\begin{equation}
D_z \frac{d^2 A_{m,im}(z|x)}{dz^2} + 1 = 0
\label{eq:immobileMAEq}
\end{equation}
By applying the following boundary conditions:
\begin{subequations}\label{eq:mobimmobMABCs}
\begin{equation}
A_{m,im} (z = 0|x) = A_{m,m} (x)
\label{eq:mobimmobMABC1}
\end{equation}
\begin{equation}
\left. {- D_z \frac{d A_{m,im}(z|x)}{dz}} \right|_{z = L}  = 0
\label{eq:mobimmobMABC2}
\end{equation}
\begin{equation}
A_{m,m}(x = 0) = 0
\label{eq:mobimmobMABC3}
\end{equation}
\end{subequations}
The solutions of~(\ref{eq:mobileMAEq}) and~(\ref{eq:immobileMAEq}) are given as
\begin{subequations}\label{eq:mobimmobMASol}
\begin{equation}
A_{m,m}(x) = \frac{x}{v_x} \left( 1 + \frac{n_{im} L}{n_m l} \right)
\label{eq:mobimmobMASol1}
\end{equation}
\begin{equation}
A_{m,im}(z|x) = \frac{1}{2 D_z}z(2L - z) + A_{m,m}(x)
\label{eq:mobimmobMASol2}
\end{equation}
\end{subequations}
Eq.~(\ref{eq:mobimmobMASol1}) corresponds to the result of~\cite{Bethke2002}, the so-called groundwater age paradox, which is actually a simple direct consequence of the hypotheses of steady-state conditions, constant porosity and no diffusion in the direction of aquifer flow in the aquitard (\cite{Ginn2009}). Clearly, the mean age in the aquifer is a function of the ratio of the water volumes of the two systems ($\frac{n_{im} L}{n_m l}$) to the kinematic age. Eq.~(\ref{eq:mobimmobMASol2}) indicates that the age in the aquitard is a travel time by pure diffusion from the aquifer to the aquitard midpoint. It is parabolic with its maximum located at the midpoint ($\max [A_{m,im} ] = A_{m,m} + L^2 /2D_z$, when $z = L$). Fig.~\ref{fig:fig4} illustrates the distributions of the groundwater mean age for the same groundwater flow system in Fig.~\ref{fig:fig3}. Fig.~\ref{fig:fig5} illustrates the distributions of the kinematic, apparent, and mean ages along the flow direction in the mobile region and the apparent and mean ages across the immobile region (at $x = 1000$m) of the two-dimensional system shown in Figures~\ref{fig:fig3} and~\ref{fig:fig4}. It is clear from Fig.~\ref{fig:fig5} that the apparent age overestimates the kinematic age but underestimates the groundwater mean age in the aquifer. Groundwater mean age in the aquitard is significantly higher than the apparent age although both ages are influenced by the mixing processes in the aquitard. This is because the apparent age is influenced by the decay rate of the solute. From Eq.~(\ref{eq:mobileAA}), although the relationship between apparent age and mean age is more complex than in the one-dimensional system, it is unique and thus the apparent age can be used to estimate the mean age for a given distance from the source, groundwater velocity, the rate of decay for the solute, and the thickness and diffusive properties of the immobile region surrounding the mobile region.

\subsection{Two-Dimensional Analysis}
Effects of transverse mixing can be analyzed by considering a two-dimensional system where the flow velocity can be approximated by a one-dimensional uniform velocity ($v_x$) but where recharge is limited at the origin ($x = 0$, $y = 0$). The kinematic age is approximated along the streamline from the source ($y = 0$) using the location and the one-dimensional velocity:
\begin{equation}
A_k(x, y = 0) \approx \frac{x}{v_x}
\label{eq:kinematicAge2D}
\end{equation}

The steady-state transport of a decaying solute is described by
\begin{equation}
- v_x \frac{\partial C(x,y)}{\partial x} + \nabla \cdot {\bf{D}} \cdot \nabla C(x,y) - \lambda C(x,y) = 0
\label{eq:solute2DADE}
\end{equation}
where the two-dimensional operator is defined as $\nabla  \equiv (\partial /\partial x,\partial /\partial y)$. When the dispersion coefficient ${\bf{D}} \equiv (D_{xx}, D_{xy}; D_{yx}, D_{yy})$ is such that $D_{xx} = \alpha_l v_x + D_{diff}$, $D_{xy} = D_{yx} = 0$, and $D_{yy} = \alpha_t v_x + D_{diff}$, Eq.~(\ref{eq:solute2DADE}) simplifies in
\begin{equation}
- v_x \frac{\partial C(x,y)}{\partial x} + D_{xx} \frac{\partial^2 C(x,y)}{\partial x^2} + D_{yy} \frac{\partial^2 C(x,y)}{\partial y^2} - \lambda C(x,y) = 0
\label{eq:solute2DADE2}
\end{equation}
For the given boundary condition of continuous mass injection with the rate of $Q C_0$ ($\textrm{M} \textrm{T}^{-1}$) at the origin, where $Q$ is the inflow rate and $C_0$ the flux concentration, the solution of Eq.~(\ref{eq:solute2DADE2}) is given by
\begin{equation}
C(x,y) = \frac{Q C_0}{2\pi n\sqrt {D_{xx} D_{yy}}}\exp \left( {\frac{v_x x}{2D_{xx}}} \right) K_0 \left( {\sqrt {\left( {\frac{v_x^2}{4D_{xx}} + \lambda } \right)\left( {\frac{x^2}{D_{xx}} + \frac{y^2}{D_{yy}}} \right)} } \right)
\label{eq:solute2Dsol}
\end{equation}
where $n$ is the porosity and $K_0$ is the modified Bessel function of second kind and zero order. When an asymptotic form of $K_0(p)$ ($\approx \sqrt {\pi / 2 p}e^{- p}$ for large $p$) is introduced in Eq.~(\ref{eq:solute2Dsol}) (e.g. see~\cite{Wilson1978}), then it reads:
\begin{equation}
C(x,y) = \frac{Q C_0}{n \sqrt {8\pi D_{xx} D_{yy} r/B}}\exp \left( \frac{x - r}B \right)
\label{eq:solute2Dsol2}
\end{equation}
where
\begin{equation}
B = \frac{2D_{xx}}{v_x} \quad\quad \textrm{and} \quad\quad r = \sqrt {\left( 1 + \frac{2B\lambda}{v_x} \right)\left( x^2 + \frac{D_{xx}}{D_{yy}}y^2 \right)}
\label{eq:Bandrdef}
\end{equation}
If the apparent age is defined from the ratio in concentrations of decaying and conservative solutes ($\lambda = 0$), then it is given as
\begin{subequations}
\begin{equation}
A_a(x,y) = - \frac{1}{\lambda} \ln \left( {\frac{C}{{C(\lambda = 0)}}} \right) = \frac{1}{\lambda}\left( {\frac{1}{4} \ln \kappa - \frac{r}{B}\left( {\frac{{1 - \sqrt \kappa}}{{\sqrt \kappa}}} \right)} \right)
\label{eq:apparentage2D}
\end{equation}
\begin{equation}
A_a(x,y = 0) = \frac{1}{\lambda}\left( {\frac{1}{4}\ln \kappa - \frac{x}{B}\left( {1 - \sqrt \kappa} \right)} \right)
\label{eq:apparentage2Dyzero}
\end{equation}
\end{subequations}
with
\begin{equation}
\kappa = 1 + \frac{{2B\lambda }}{{v_x}}
\label{eq:kappadef}
\end{equation}
An asymptotic solution for $A_m(x,y)$ is obtained when solving the mean age equation
\begin{equation}
- v_x \frac{\partial A_m(x,y)}{\partial x} + D_{xx} \frac{\partial^2 A_m(x,y)}{\partial x^2} + D_{yy} \frac{\partial^2 A_m(x,y)}{\partial y^2} + 1 = 0
\label{eq:MAeqn2D}
\end{equation}
Using the zero mean age boundary condition $A_m(x = 0, y = 0) = 0$, this solution is:
\begin{subequations}
\begin{equation}
A_m(x,y) = \frac{1}{v_x} \left( {\sqrt {x^2 + \frac{D_{xx}}{D_{yy}}y^2} + \frac{D_{xx}}{v_x}} \right)
\label{eq:MAsol2D}
\end{equation}
\begin{equation}
A_m(x,y = 0) = \frac{x}{v_x} + \frac{D_{xx}}{v_x^2}
\label{eq:MAsol2Dyzero}
\end{equation}
\end{subequations}
Eqs.~(\ref{eq:apparentage2D}) and~(\ref{eq:MAsol2D}) show a dependency of age with transverse dispersion $D_{yy}$. The ratio of apparent age and mean age is more complex than in the one-dimensional system but it still can be computed and potentially used to assess the dispersive properties of the aquifer. Eqs.~(\ref{eq:apparentage2Dyzero}) and~(\ref{eq:MAsol2Dyzero}) indicate that the apparent and mean ages are not influenced by the transverse dispersion along the streamline from the source. The apparent and mean ages, compared to the kinematic age, yield the following relation:
\begin{subequations}
\begin{equation}
r_{A_{mk}}(x,y = 0) = \frac{A_m}{A_k} = 1 + \frac{D_{xx}}{x v_x}
\label{eq:ratioA2D1}
\end{equation}
\begin{equation}
r_{A_{ak}}(x,y = 0) = \frac{A_a}{A_k} = \frac{v_x}{4\lambda x}\ln \kappa + \frac{v_x^2}{2\lambda D_{xx}}\left( {\sqrt \kappa - 1} \right)
\label{eq:ratioA2D2}
\end{equation}
\end{subequations}
In Eq.~(\ref{eq:ratioA2D1}), it is clear that ratio between the kinematic and mean ages becomes unity for large $x$ and $v_x$, and small $D_{xx}$. When $D_{xx} \approx \alpha_l v_x$, $r_{A_{mk}} \approx 1 + \alpha_l / x$ and thus the mean and kinematic ages becomes similar as the location is farther from the source compared to the longitudinal dispersivity. In Eq.~(\ref{eq:ratioA2D2}), it is also clear that $r_{A_{ak}} \approx \frac{v_x^2}{2\lambda D_{xx}}\left( {\sqrt \kappa - 1} \right)$ for large $x$, in which case the one-dimensional solution~(\ref{eq:ratios1}) is recovered. As a result, it can be concluded that the effect of dimensionality on age bias is minor if the velocity field is uniform.

\section{Implications for Tritium Age Dating}
Transient transport of isotopic signatures has been widely used to measure the age of groundwater. One of the most popular radioisotopes in dating modern groundwater is the tritium which was produced by the atmospheric testing of thermonuclear bombs between 1951 and 1980 (activity peaked at 1963, e.g. see~\cite{Maloszewski1982},~\cite{Zuber2005},~\cite{IAEAUNESCO2001},~\cite{Morgenstern2010}). There can be various different approaches to dating groundwater using tritium as explained by~\cite{Clark1997} and they may be categorized as: (\emph{i}) identifying the spatial location of concentration (or activity) peak, (\emph{ii}) calculating the time for decay from a known input level to the measured level, (\emph{iii}) analyzing concentration time series at a specific point. \cite{Clark1997} indicated that the tritium age dating could be most appropriate when the influence of the mixing and dispersion is minimal for isotope activities. It is interesting to note that the spatial location of the tritium activity peak in the subsurface is influenced only by the advection and thus it is related only to the kinematic age, while the attenuation and decay can be strongly influenced by the mixing or other dilution processes. Both of these measures, however, could be misleading if they are used to analyze the mean residence time or mean age of the sample.\\

Transport of tritium in a one-dimensional domain can be described by the following advection-dispersion-reaction equation:
\begin{equation}
\frac{\partial C(x,t)}{\partial t} =  - v\frac{\partial C(x,t)}{\partial x} + D\frac{\partial ^2 C(x,t)}{\partial x^2} - \lambda C(x,t)
\label{eq:TritiumADE}
\end{equation}
For the simplistic analysis, a steady flow field and an initial Gaussian slug input of tritium is assumed at the origin:
\begin{equation}
C(x, t = 0) = \frac{1}{\sqrt {4 \pi D t_0}}\exp \left( - \frac{x^2}{4 D t_0} - \lambda t_0 \right)
\label{eq:GaussianSlugTritium}
\end{equation}
The variable $t_0$ denotes the initial time at which the concentration defined in Eq.~(\ref{eq:GaussianSlugTritium}) is taken as the initial concentration. The solution of Eq.~(\ref{eq:TritiumADE}) is given by
\begin{equation}
C(x,t) = \frac{1}{\sqrt {4 \pi D(t + t_0)}}\exp \left( {- \frac{(x - vt)^2}{4 D (t + t_0)} - \lambda (t + t_0)} \right)
\label{eq:TritiumSol}
\end{equation}
In Eq.~(\ref{eq:TritiumSol}), the position of the peak is located at $x = v t$ with the peak value being
\begin{equation}
C(x = v t,t) = \frac{e^{- \lambda (t + t_0)}}{\sqrt {4 \pi D(t + t_0)}}
\label{eq:TritiumPeak}
\end{equation}
If the location of the peak ($x = x_{peak}$) is known at given time $t = t_{peak}$, groundwater velocity is calculated as $v = x_{peak} / t_{peak}$  and the kinematic age of groundwater at $x = x_{peak}$ is given by $A_k (x = x_{peak} ) = t_{peak}$, which is same as the mean age of groundwater in a one-dimensional flow system ($A_k = A_m$). It is noted again that this is based only on the fact that the peak migrates with the rate of groundwater velocity and is not related to the decay of the solute.\\

If the age is calculated from the change in peak concentration over the time period $t_{peak}$ (also known as time lag method), the apparent age of groundwater is given by
\begin{align}\label{eq:weakform32}
A_a(x = x_{peak}) = {} & - \frac{1}{\lambda}\ln \left( \frac{C}{C_0} \right) =
- \frac{1}{\lambda} \ln \left( {\frac{\sqrt {t_0} e^{- \lambda t_{peak}}}{\sqrt {t_{peak} + t_0}}} \right) \\
 & = t_{peak} + \frac{1}{\lambda} \ln \left( \sqrt {1 + \frac{t_{peak}}{t_0}} \right) \nonumber
\label{eq:AA_tritium}
\end{align}
Because the apparent age calculation is based on the rate of decrease in concentration values, additional dilution by the dispersion or mixing can make it greater than the kinematic age. It is clear that apparent age increases as the time for the chemical measure increases.\\

\begin{figure}[h]
  \begin{center}
  \includegraphics[width=\textwidth]{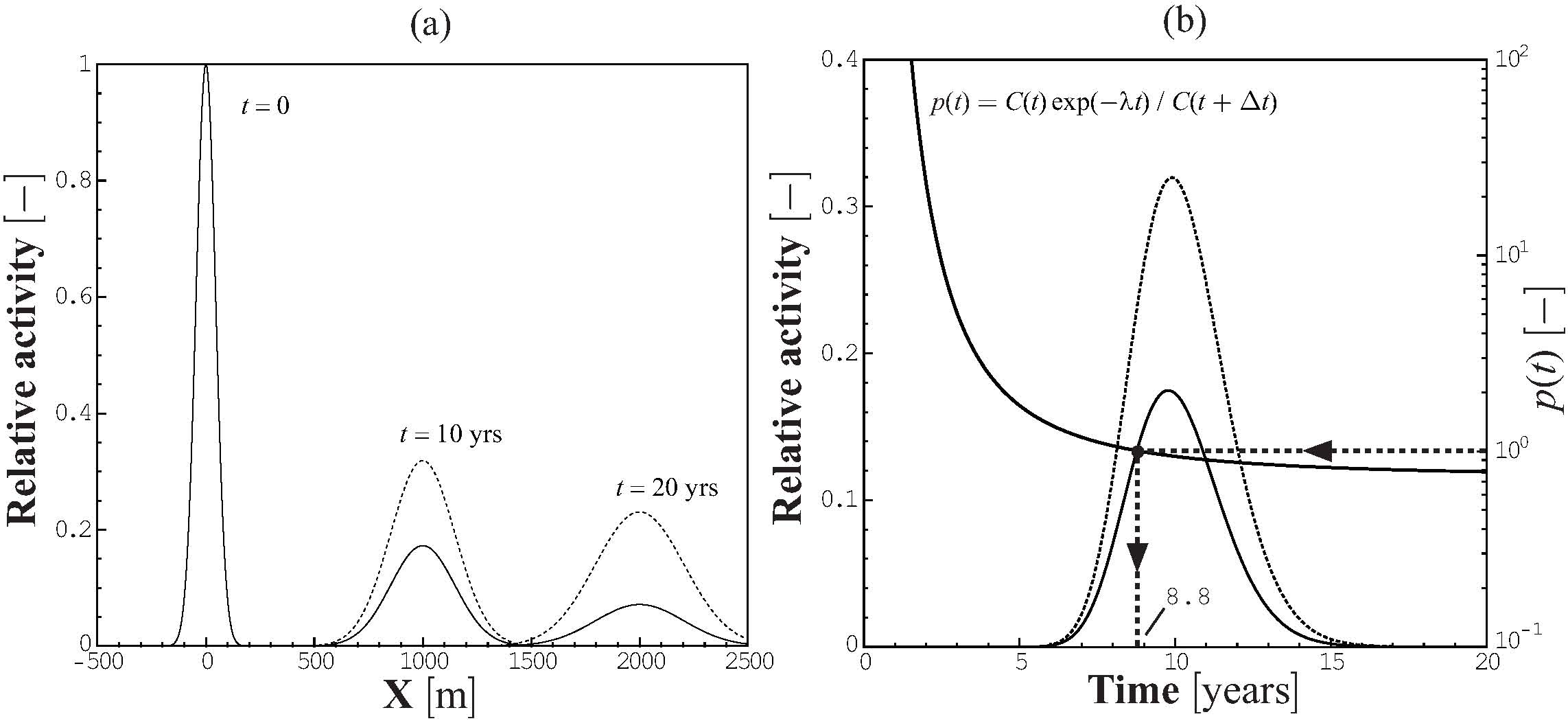}
  \end{center}
  \caption{\small (a): Spatial distribution of tritium activity; (b): Temporal distribution of tritium activity, with indication on the criterion $p(t)$ of~\cite{Fritz1991}. Solid lines: decaying tracer; Dashed line: conservative tracer.}
 \label{fig:fig6}
\end{figure}

In a hypothetical one-dimensional flow field, groundwater velocity ($v$) is assumed to be 100 m/yr with the dispersion coefficient of 1000m. For the sake of simplicity, activity at time zero is assumed to be Gaussian with the peak at $x = 0$ ($t_0  = 1{\rm{yr}}$). Fig.~\ref{fig:fig6}a shows the spatial distribution of tritium activity at $t =$~0, 10, and 20 years and Fig.~\ref{fig:fig6}b shows the yearly time series at $x = 1000$m. The spatial distribution of tritium activity at $t = 10$years indicates that it peaks at $x = 1000$m and thus the kinematic age or the mean age can be calculated as 10 years. This is valid also for the conservative tracer. If the simple rate of decay is used to calculate the apparent age at $t = 10$years, then $A_a(x = x_{peak} ) \approx 31.5$. Using the ratio between non-decaying and decaying solutes, one obtains $A_a(x = x_{peak}) = - (1/\lambda) \ln \left( C/C_{\lambda = 0} \right) \approx 11.0$years.\\

\cite{Fritz1991} estimated groundwater mean residence time using tritium activities measured from the samples collected at the same location in different years. They suggested to use the criterion $p(t) = C(t) \exp (- \lambda t)/C(t + \Delta t)$ to determine whether the peak passed the location or not. If $p(t) > 1$, then the peak has passed the observation point, whereas if $p(t) < 1$, then the observation point is sampling the leading edge. From Fig.~\ref{fig:fig6}b, $p(t \approx 8.8 \, {\rm{years}}) = 1$ and so the age is determined to be 8.8 years at $x = 1000$ m. The method by~\cite{Fritz1991} is derived from the decay equation. If the observation point is sampling the leading edge, then $dC(t)/dt > - \lambda C(t)$ and for the tail, $dC(t)/dt < - \lambda C(t)$ since the advection and dispersion increases the concentration before the peak passes the location while concentration decreases after the peak. Although this is a semi-quantitative approach to determine the relative location of the peak in the flow system, it closely approximates the kinematic. \cite{Delhez2008} however show how such apparent ages systematically underestimate the mean age.\\

It is implied from the above analyses that the isotopic age dating could provide an approximate mean age or kinematic age when the flow system is homogeneous and uniform (or approximated as one-dimensional), when the mixing is minimal, and if the spatial and temporal location of the peak is used to measure the age. However, the calculated age is based on the radioactivity (radiometric age) and may not be representative for the mean residence time of the sample or groundwater velocity at given locations.

\section{Numerical Example}
\begin{figure}[!t]
  \begin{center}
  \includegraphics[width=0.8\textwidth]{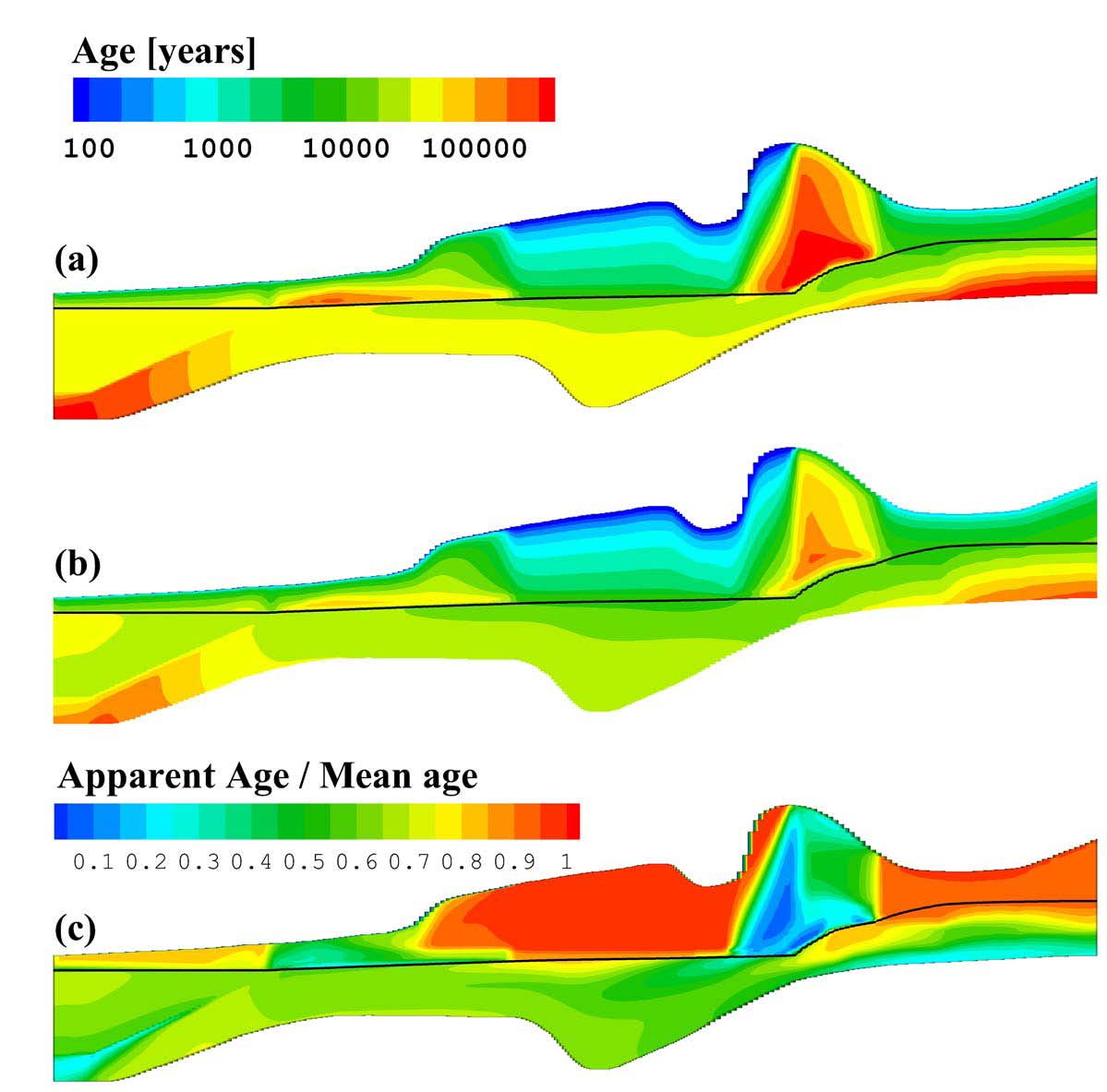}
  \end{center}
  \caption{\small (a) Mean and (b) radiometric ages simulated using Eqs.~(\ref{eq:apparentAge}) and~(\ref{eq:1DADE}) in Yucca Mountain, Nevada. (c) Relative difference. Steady flow field is assumed for both simulations and the black solid lines indicate the location of the water table.}
  \label{fig:fig7}
\end{figure}

\cite{Schwartz2010} analyzed the $^{14}\textrm{C}$ apparent age at Yucca Mountain in Nevada, in a steady flow system and also in a historic transient since the last glacial maximum, and indicated that accounting for the historic transients in recharge in the flow system makes a significant difference in apparent age distribution at present compared to that simulated using the present-day steady flow system. In this study, the mean age simulated using Eq.~(\ref{eq:apparentAge}) is compared to the radiometric age translated from the results of the simulation using Eq.~(\ref{eq:1DADE}) in the same steady flow system used by~\cite{Schwartz2010} (Fig.~\ref{fig:fig7}). Details for the domain, initial and boundary conditions, and all the parameters used for the simulations are referred to~\cite{Schwartz2010}: in the two-dimensional vertical flow domain (approximately 20 km long and 2 km deep) consisting of different volcanic aquifers and aquitards, water table location is 500 m below the ground surface or deeper, and recharge ranges between 0.05 to 10 mm/year.\\

Fig.~\ref{fig:fig7} shows the distributions of the simulated mean and radiometric ages in the steady flow system. From the results in Fig.~\ref{fig:fig7}, it is clear that the simulated ages are similar in pattern (if the water is older in terms of radiometric age, then the mean residence time is greater based on the mean age) but the radiometric age underestimates the mean residence time, especially when the water becomes older, agreeing with the analysis in the previous sections. In the unsaturated zone above the water table where transport is mostly advection-dominated, the two age distributions are almost identical except for the low permeable-low recharge region where the Caldera Barrier exists (\cite{Schwartz2010}). In the saturated zone, the two age distributions are relatively distinct since the mixing of waters of different origins and the dispersion processes play more important roles.

\section{Conclusions}
The presented one-dimensional and two-dimensional analyses reveal that the differences between mean age (as obtained from the direct physical modelling of mean age fate) and apparent age (as obtained from isotope concentrations) can be very important, and that the competition between decay and dispersion coefficient is a major factor that can explain such differences. More precisely, the larger the product $\lambda D$, the larger the difference between mean and apparent ages. It is also shown that apparent age systematically overestimates kinematic age and underestimates mean age. This statement is verified in the context of aquifer-aquitard systems, for which coupled one-dimensional and two-dimensional solutions also show that mean age in the aquitard is significantly higher than the apparent age although both ages are influenced by the mixing processes in the aquitard. A correction to the average apparent velocity $v_a$ that is obtained from two apparent age dates at two distinct locations is also provided. This correction is a direct function of the dimensionless variable $\frac{v_a^2}{\lambda D}$, and indicates that the average velocity may be calculated from the apparent velocity only when $v_a > \sqrt{\lambda D}$. The analytical analysis is concluded by the study of the effects of transverse mixing on apparent and mean ages. The results indicate a relatively minor effect of dimensionality for uniform flow fields, and that apparent and mean ages are not influenced by transverse dispersion along the streamline from the source.\\
The differences in apparent age and mean age observed with simple system configurations can be expected to be much larger in more realistic and complex situations, e.g. when several flow systems tend to mix, with the consequence that dispersion of ages is not the result of mechanical dispersion only, but also a consequence of the travel
times repartition within each flow system (being itself a consequence of the shape and
length of flow paths, thus of the geological structure of the aquifer and of the boundary
conditions). A numerical analysis of age distribution at Yucca Mountain in Nevada is presented to illustrate the magnitude of these age differences one can expect in realistic situations.\\
Our results suggest that the use of the classical piston-flow model (and by extension of the exponential model
and others) tends to be irrelevant for age dating and particularly for the interpretations that are frequently carried out to derive average flow velocities. Age definitions based on the radioactivity of isotopes may not be representative of the mean residence time of the sample or for the groundwater velocity at given locations, and may not be suitable in many cases for the calibration of hydrogeological models.

\section*{Acknowledgements}
Eric Deleersnijder is a Research Associate with the Belgian Fund for Scientific Research (F.R.S.-FNRS). His contribution to the present study was achieved in the framework of the Interuniversity Attraction Pole TIMOTHY (www.climate.be/timothy), which is funded by the Belgian Science Policy office (www.belspo.be) under contract IAP6.13.

\bibliographystyle{elsarticle-harv}
\newpage
\bibliography{BiblioA1}

\end{document}